\documentclass{pasa}
\usepackage[flushleft]{threeparttable}
\usepackage{graphicx}
\usepackage[caption=false]{subfig}
\usepackage{url}

\begin{document}

\title[A GPU-Based Wide-Band Radio Spectrometer]{A GPU-Based Wide-Band Radio Spectrometer}
\author[Chennamangalam et al.]{Jayanth Chennamangalam$^{1,7}$, Simon Scott$^2$,
    Glenn Jones$^3$, Hong Chen$^2$, John Ford$^5$, Amanda Kepley$^5$,\\
    D. R. Lorimer$^{1,5}$, Jun Nie$^4$, Richard Prestage$^5$, D. Anish Roshi$^6$,
    Mark Wagner$^2$, \and Dan Werthimer$^2$\\
\affil{$^1$Department of Physics and Astronomy, West Virginia University,
       PO Box 6315, Morgantown, WV 26506, USA}
\affil{$^2$University of California, Berkeley, CA 94720, USA}
\affil{$^3$Columbia University, New York, NY 10027, USA}
\affil{$^4$Xinjiang Astronomical Observatory, Chinese Academy of Sciences, Urumqi 830011, China}
\affil{$^5$NRAO, Green Bank Observatory, PO Box 2, Green Bank, WV 24944, USA}
\affil{$^6$NRAO, Charlottesville, VA, 22903, USA}
\affil{$^7$Corresponding author. Present address: Astrophysics, University of
Oxford, Denys Wilkinson Building, Keble Road, Oxford OX1 3RH, UK.
Email: jayanth@astro.ox.ac.uk}}

\begin{abstract}
The Graphics Processing Unit (GPU) has become an integral part of astronomical
instrumentation, enabling high-performance online data reduction and
accelerated online signal processing. In this paper, we describe a wide-band
reconfigurable spectrometer built using an off-the-shelf GPU card. This
spectrometer, when configured as a polyphase filter bank (PFB), supports a
dual-polarization bandwidth of up to 1.1 GHz (or a single-polarization
bandwidth of up to 2.2 GHz) on the latest generation of GPUs. On the other
hand, when configured as a direct FFT, the spectrometer supports a
dual-polarization bandwidth of up to 1.4 GHz (or a single-polarization
bandwidth of up to 2.8 GHz).
\end{abstract}


\maketitle

\section{Introduction}

\noindent Astronomical data acquisition and online reduction of data are
steadily becoming more resource-intensive, not just for new and upcoming
telescopes such as the Low Frequency Array (LOFAR) and the Square
Kilometre Array (SKA), but also for new instruments at established facilities.
Field Programmable Gate Arrays (FPGAs) have long been used at the output of
Analog-to-Digital Converters (ADCs) for data reduction and/or packetization,
followed by a computer that manages the recording of data to disk.
FPGAs have traditionally been considered suitable for high-bandwidth
applications, but the relative difficulty in programming them and the lack of
support for floating-point arithmetic, coupled with the relatively inexpensive
pricing of Graphics Processing Unit (GPU) cards, have popularized the use of
GPUs in astronomical instrumentation. Several real-time GPU-based signal
processing systems intended for pulsar astronomy have been developed in recent
years \cite[for instance]{ran09,mag11,arm12,bar12,mag13}. Many new
instruments combine the high-bandwidth data acquisition capability of FPGAs
with the high-performance data reduction capability of GPUs, glueing them
together with high-throughput networking hardware. Such a heterogeneous
architecture is expected to scale up to meet the data-handling requirements of
future instruments and telescopes.

In this paper, we give an overview of a heterogeneous, wide-bandwidth,
multi-beam spectrometer that we have built for the Green Bank Telescope (GBT),
focussing on the GPU-based spectrometry code and its performance. This
spectrometer forms part of the `Versatile GBT Astronomical Spectrometer'
(VEGAS) \cite{ros11,for13}. VEGAS has multiple modes of operation that are broadly
classified into two categories -- the so-called high bandwidth (HBW) and
low-bandwidth (LBW) modes. The HBW modes are characterized by
higher bandwidths and faster spectral dump rates. The HBW mode spectrometry
takes place exclusively on Field-Programmable Gate Array (FPGA)
boards\footnote{The Reconfigurable Open Architecture Computing Hardware II
(ROACH II) platform.}, with integrated spectra sent to data recording PCs. The LBW modes,
on the other hand, involve heterogeneous instrumentation, combing FPGAs for
data acquisition followed by GPUs for spectrometry. These modes are based on
GPUs as they require a larger number of channels (up to 524288) than what FPGAs
can support. The LBW modes have slower spectral dump rates (i.e.,
more spectra are integrated) compared to the HBW modes. Although these are
`low-bandwidth' modes in the context of VEGAS, these modes are based on code
that, with sufficient spectral integration time, can support a
per-polarization bandwidth of up to 1.4 GHz. Since these modes make use of both
FPGAs and GPUs, for the remainder of this paper, they will be referred to as
`heterogeneous modes'. 

The organization of this paper is as follows. In \S\ref{sec_gpu}, we introduce
the GPU-programming paradigm, in \S\ref{sec_vegas_desc} we describe our
software, and in \S\ref{sec_perf} we explain our benchmarking procedure and
performance figures.

\section{The GPU-Programming Paradigm} \label{sec_gpu}

In the past, computing performance was improved most commonly by using smaller
silicon features and increasing the clock rate. Since computer designers are no
longer able to increase the clock rate further due to power constraints,
parallelization is the primary method to improve performance in recent times.
The Central Processing Unit (CPU) of a typical personal computer (PC) has
traditionally contained a single instruction-processing core that can perform
only one operation at a time. Multi-tasking on a PC powered by such a CPU is
usually achieved by interleaving tasks in time. This obviously degrades the
performance of time-critical tasks such as rendering graphics for computer
games. One solution to this problem is to offload graphics processing to a
dedicated co-processor, the GPU. The GPU contains multiple processing cores
that enables it to run multiple instructions simultaneously\footnote{Even
though modern CPUs contain multiple processing cores (on the order of tens of
cores), modern GPUs far surpass them, having cores on the order of hundreds to
thousands.}. This parallelization makes it suitable not just for graphics
processing, but also for general purpose computing that requires high
performance. Modern GPUs are designed with this in mind, and programming
platforms are available that let developers take advantage of this computing
power.

The most common general purpose GPU programming platform is Compute Unified
Device
Architecture\footnote{\url{http://www.nvidia.com/object/cuda_home_new.html}}
(CUDA). CUDA lets developers access the hardware (the parallel compute engine
in the GPU) using programming instructions. This is enabled by extending the C
language to invoke routines that run on the GPU and using CUDA libraries for
numerical computation and signal processing, such as CUDA Basic Linear Algebra
Subroutines (CUBLAS) and CUDA Fast Fourier Transform (CUFFT).

Since GPUs are suitable computing platforms for data-parallel applications,
they are increasingly used as dedicated co-processors for data analysis
applications that use the high-performance hardware to accelerate their
time-critical paths. This also makes GPUs ideal for data-acquisition
instruments such as VEGAS.

\section{Overview of VEGAS} \label{sec_vegas_desc}

The heterogeneous modes of operation of
VEGAS\footnote{\url{http://www.gb.nrao.edu/vegas/}} and their specifications
are given in Table~\ref{tab_modes}. These modes are divided into
single-sub-band modes and eight-sub-band modes. The single-sub-band modes can have
32768 to 524288 channels, with sub-band bandwidths in the range 11.72 MHz to
187.5 MHz, whereas the eight-sub-band modes have 4096 to 65536 channels with
bandwidths ranging from 15.625 MHz to 23.44 MHz.

\begin{table*}
\begin{center}
\begin{tabular}{ccccc}
\hline
\parbox[b]{1.5cm}{Number of sub-bands per pol.}
& \parbox[b]{1.5cm}{Sub-band bandwidth}
& \parbox[b]{1.5cm}{Number of channels per sub-band\\per pol.}
& \parbox[b]{1.5cm}{Spectral resolution}
& \parbox[b]{1.5cm}{Min. integration time}\\
& (MHz) &  & (KHz) & (ms)\\
\hline
1 & 100.0 -- 187.5 & 32768 -- 131072 & \hphantom{0}0.8 -- 5.7 & 10 -- \hphantom{0}30\\
1 & \hphantom{0}11.72 -- \hphantom{0}23.44 & 32768 -- 524288 & 0.02 -- 0.7 & \hphantom{0}5 -- \hphantom{0}75\\
8 & 15.625 -- \hphantom{0}23.44 & \hphantom{0}4096 -- \hphantom{0}65536 & 0.24 -- 5.7 & \hphantom{0}5 -- 100\\
\hline
\end{tabular}
\end{center}
\caption{Heterogeneous modes of operation of VEGAS and their specifications.}
\label{tab_modes}
\end{table*}

\begin{figure*}
\begin{center}
\includegraphics[scale=0.8]{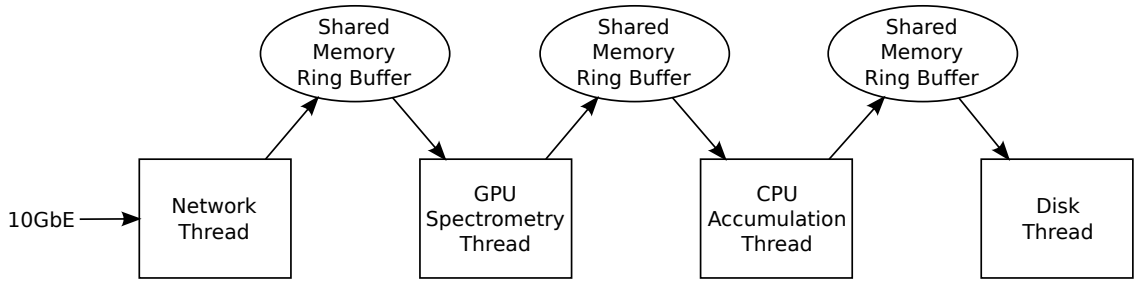}
\caption{Data flow diagram of the software part of the VEGAS data acquisition
pipeline. This paper focuses on the software used in the GPU spectrometry
thread.}
\label{fig_block}
\end{center}
\end{figure*}

Figure~\ref{fig_block} shows a block diagram of the software
section of the VEGAS heterogeneous-mode
data acquisition pipeline. In the heterogeneous modes, the FPGA board
packetizes the signal sampled by an ADC and sends it over 10-Gigabit Ethernet
(10GbE) to a PC with a GPU. The VEGAS software pipeline, based on the
Green-Bank Ultimate Pulsar Processing Instrument
(GUPPI; Ransom et al. \shortcite{ran09}) is made up of multiple concurrent
threads, each associated with a separate CPU core. The first thread, called the
`network thread' reads packets off the network and writes the payload to a
shared memory ring buffer. The next thread, called the `GPU thread' reads the
data off the buffer and performs spectrometry, including accumulation of
multiple spectra, if needed. Once the accumulated spectra are ready, the output
is written to another ring buffer from which the third thread -- the `CPU
thread' -- reads data and performs further accumulation as needed. Once this is
done, the output is sent to the `disk thread' that writes it to disk. This
paper describes the spectrometer implemented in the GPU thread.

VEGAS supports multi-beam receivers, in which case the signal
from each beam is processed by a separate software data acquisition pipeline.
The implementation utilizes dual-socket, dual-NIC, dual-GPU PCs, wherein
one PC processes signals from two beams independently.

\section{The GPU Spectrometer}

Spectrometry is a Discrete Fourier Transform (DFT; see, for example,
Bracewell \shortcite{bra99}) operation, usually implemented as a Fast Fourier Transform
(FFT) for its performance benefits. Due to the finite length of the `DFT
window' (the number of input time samples), the single-bin frequency domain
response of the DFT is not rectangular, but is a sinc function, with side lobes
spread across the entire bandwidth. This `spectral leakage', and the related
phenomenon of `scalloping loss' -- due to the non-flat nature of the main lobe
of the sinc function -- can be mitigated by suppressing the side-lobes of the
sinc function and changing the single-bin frequency response of the DFT to
approximate a rectangular function. One way of achieving this is
using the polyphase filter bank technique (PFB), also known as weighted
overlap-add method, in which a `pre-filter' is introduced preceding the FFT
stage (for details, see Crochiere \& Rabiner \shortcite{cro83} and Harris \&
Haines \shortcite{har11}). The GPU spectrometer described in this paper
implements an 8-tap polyphase filter bank.

The input data to our PFB spectrometer is made up of dual-polarization, 8-bit,
complex-valued samples, while the output contains $X^2$, $Y^2$, Re($XY^*$), and
Im($XY^*$), where $X$ is the Fourier transform of the horizontal polarization,
$Y$ is the Fourier transform of the vertical polarization, and $X^*$ and $Y^*$
are the corresponding complex conjugates. Note that full-Stokes spectra can
easily be derived from these values.

The high-level algorithm of the spectrometer\footnote{The GPU spectrometer code
that we have developed is available freely for download from
\url{https://github.com/jayanthc/grating}.} is as follows. Here, following
GPU-programming parlance, `host' indicates code and memory that is associated
with the CPU, whereas `device' indicates code and memory that is associated
with the GPU.

\begin{enumerate}
  \item Initialization
  \begin{enumerate}
    \item Load filter coefficients from host to device memory
    \item Create FFT plan to perform two FFTs in parallel in the case of
          single-sub-band modes and 16 FFTs in parallel in the case of
          eight-sub-band modes
  \end{enumerate}
  \item \label{loop} Copy time series data for one set of parallel FFTs to
        device
  \item Perform pre-filtering
  \item Perform parallel FFTs
  \item Accumulate spectra for desired duration
  \item Copy output to host
  \item Repeat from Step~\ref{loop}
\end{enumerate}

\subsection{Test Observations}

\begin{figure*}
\begin{center}
\includegraphics[scale=0.6]{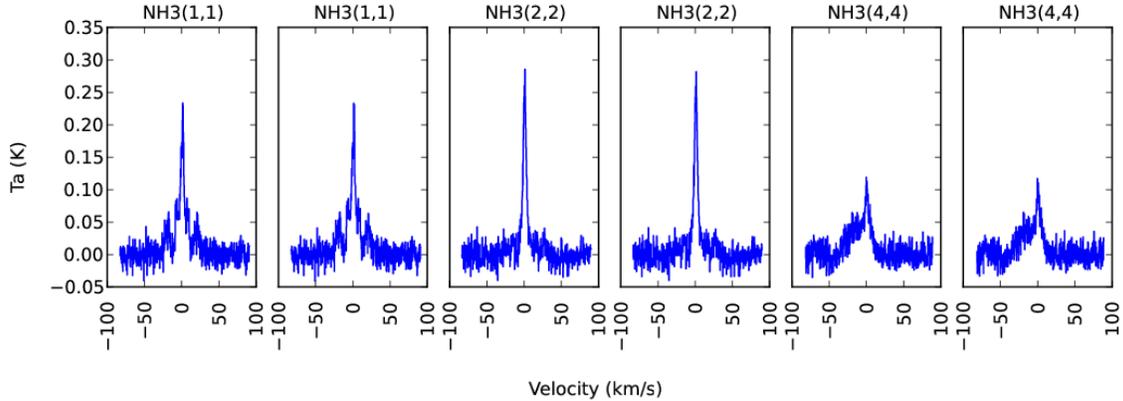}
\caption{A plot of antenna temperature versus velocity for multiple sub-bands
corresponding to one of the KFPA beams of the GBT. The plots show ammonia lines
towards RA (J2000) = 02:25:40.5, dec. (J2000) =  62:06:24, in the Galactic
HII region W3. Some of the sub-bands were tuned to the same frequencies.}
\label{fig_data}
\end{center}
\end{figure*}

We observed the Galactic HII region W3 using the seven-beam $K$-Band Focal
Plane Array (KFPA) receiver of the GBT during commissioning tests.
Figure~\ref{fig_data} shows a plot of antenna temperature versus velocity for
multiple sub-bands corresponding to one of the KFPA beams, in which ammonia
lines are visible.

\section{Benchmarking and Performance Results} \label{sec_perf}

\begin{figure*}
\begin{center}
\subfloat{
\includegraphics[scale=0.4]{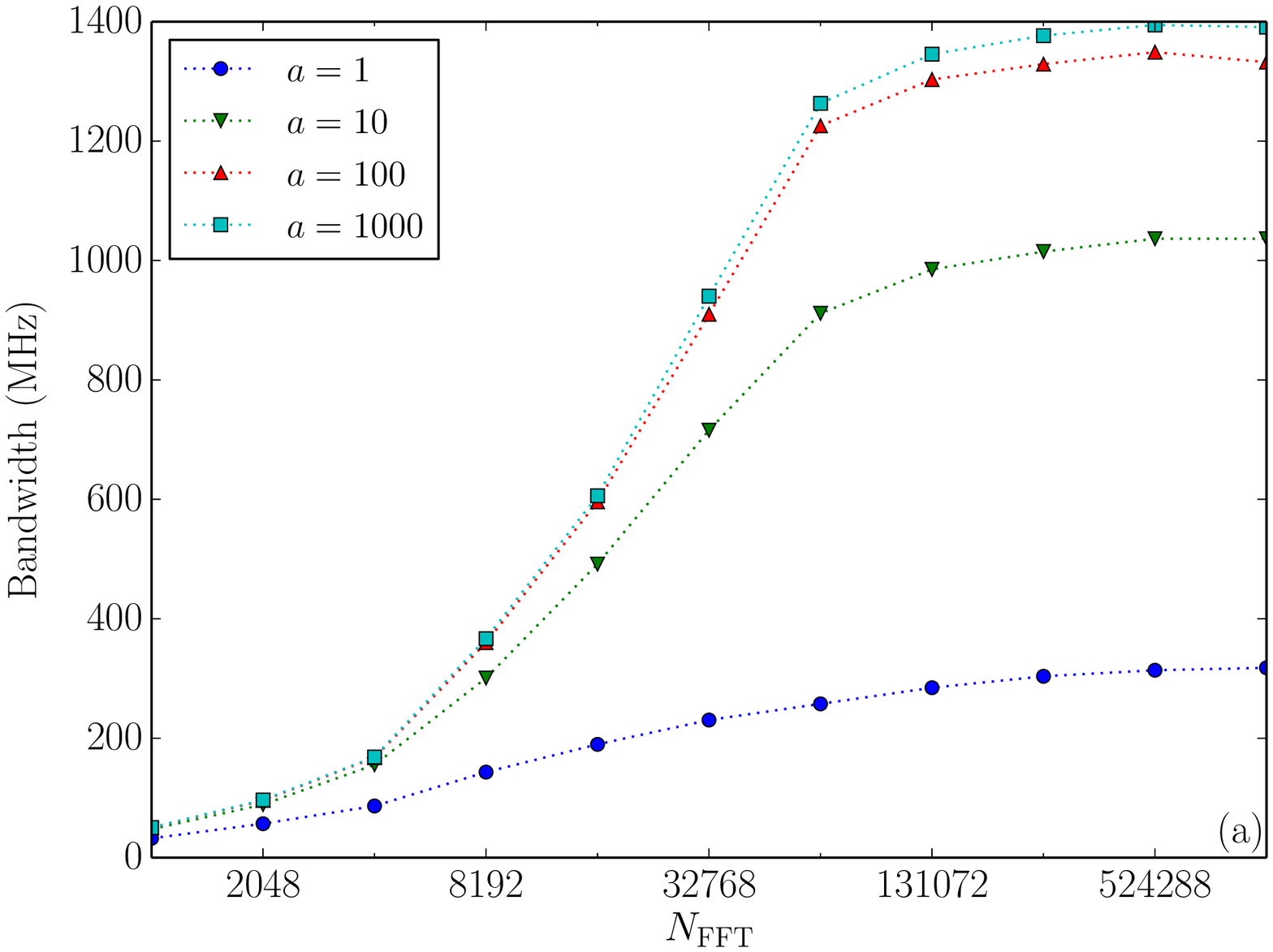}}
\subfloat{
\includegraphics[scale=0.4]{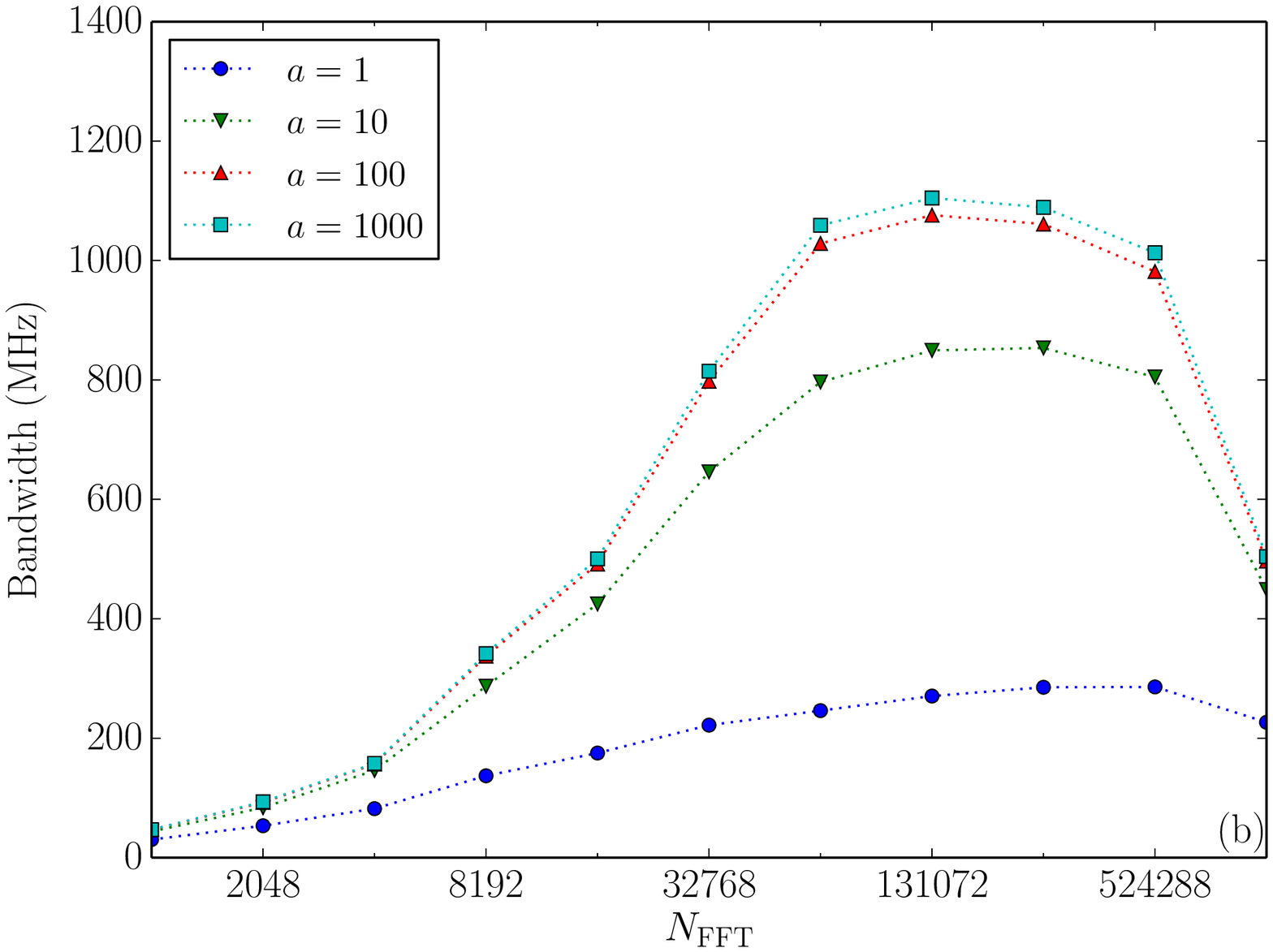}}
\caption{Bandwidth per polarization processed by the GPU spectrometer running
on an NVIDIA GeForce GTX TITAN, for (a) direct FFT and (b) 8-tap PFB, for various
values of spectral accumulation length (number of spectra accumulated), $a$.
Without the PFB technique (i.e., direct FFT), the spectrometer is able to
process a bandwidth of up to 1.4 GHz, whereas with the PFB, the maximum
bandwidth achieved is 1.1 GHz.}
\label{fig_perf}
\end{center}
\end{figure*}

Benchmarking of the software spectrometer was performed on a server-class PC
running a flavour of the Linux operating system, with an NVIDIA GeForce GTX
TITAN commercial (gaming) GPU card. A stand-alone version of the spectrometer
program was used, wherein data was read off disk files and pre-loaded in
memory, to simulate reading from the shared memory ring buffers of VEGAS
described in \S\ref{sec_vegas_desc}. Each test was repeated 100 times and we
report the average values. The peak bandwidth achieved was $\sim638$MHz
(dual-polarization), corresponding to a data rate of $\sim10.2$ Gbps which is
more than what a 10GbE link can support. The peak performance was achieved for
an FFT length of $2^{16}$, with long integrations (accumulation length of
1000 spectra). When direct FFT was used, the peak bandwidth achieved was
$\sim794$ MHz, corresponding to a data rate of $\sim12.7$ Gbps, again, more
than what is supported by 10GbE. This peak was for a $2^{20}$-point FFT with an
accumulation length of 1000 spectra. The performance of the code as a function
of transform length and accumulation length is depicted in
Figure~\ref{fig_perf}.

The performance of the code is lower at low values of FFT length
due to the following: Each FFT kernel invocation (that does either two
FFTs in parallel for single-sub-band modes, or 16 FFTs in parallel for
eight-sub-band modes) is preceded by a host-to-device data copy step and a
pre-filter stage (in the case of PFB), and followed by a device-to-host data
copy step. Given the overhead involved in launches of both the copy and
compute kernels, this translates to fewer data processing operations per unit
time, resulting in reduced performance. This becomes less of an issue at larger
FFT lengths.

\section{Conclusion}

We have developed a GPU-based PFB spectrometer that supports a
dual-polarization bandwidth of up to 1.1 GHz (or a single-polarization
bandwidth of up to 2.2 GHz). Without doing PFB (i.e., direct FFT), it supports
a dual-polarization bandwidth of up to 1.4 GHz (or a single-polarization
bandwidth of up to 2.8 GHz). This bandwidth is sufficient for most spectral
line observations, and can be traded off with spectral integration time for
some pulsar observations. Future work would involve improving the
performance of this software. The simplest way to speed it up would be to run
it on the latest generation of GPU cards. Each new generation of GPU cards
typically have, compared to its predecessors, more processing
cores and larger memory bandwidth. This naturally leads to some improvement in
performance. However, to significantly improve performance between two
consecutive generations of GPUs, the code would need to be tuned keeping in
mind the architecture of the GPU used.
A better -- albeit, brute-force -- way to speed up the code would be to
implement support for scalability, by enabling the software to take advantage
of dual-GPU cards, and/or to spread the load across multiple GPU cards. This
has the potential to increase the bandwidth processed by up to a factor of a
few, depending on the number of GPUs used. Additionally, algorithm-level and
further code-level optimizations -- such as pipelining kernel launches using
CUDA streams -- may also have the potential to yield higher performance.

\section*{Acknowledgements}
We thank Mike Clark for suggestions on code optimization, and Ben Barsdell,
Matthew Bailes, Jonathon Kocz, Gregory Desvignes, David MacMahon, and Terry
Filiba for useful discussions. We also thank the anonymous referee for comments
that served to clarify the paper. NVIDIA, GeForce, GeForce GTX TITAN, and GTX
are trademarks and/or registered trademarks of NVIDIA Corporation in the U.S.
and/or other countries.

\end{document}